\documentclass[aps,prl,twocolumn,superscriptaddress,showpacs,10pt]{revtex4-1}

\usepackage{amsmath, amsthm, amssymb}
\usepackage{bm}
\usepackage{braket}
\usepackage{array}
\usepackage{longtable}
\usepackage{color}  
\usepackage{graphicx}
\usepackage{dcolumn}
\newcolumntype{.}{D{.}{.}{-1}}
\usepackage{multirow}
\usepackage[figuresright]{rotating}
\usepackage{units}
\usepackage{subfigure}
\usepackage[latin1]{inputenc}
\usepackage{textcomp}
\usepackage{gensymb}
\usepackage[rightcaption]{sidecap}
\usepackage{nicefrac}
\usepackage{xspace}
\usepackage[version=3]{mhchem}
\usepackage{tabularx, booktabs}

\newcommand{\cacro}{Ca$_{10}$Cr$_7$O$_{28}$\xspace}

\newcommand{\Cr}{Cr$^{5+}$\xspace}
\newcommand{\Crnm}{Cr$^{6+}$\xspace}

\begin{document}

\title{Crystal growth, structure and magnetic properties of \cacro}

\author{Christian Balz}
\email{christian.balz@helmholtz-berlin.de}
\affiliation{Helmholtz-Zentrum Berlin f\"ur Materialien und Energie, 14109 Berlin, Germany}
\affiliation{Institut f\"ur Festk\"orperphysik, Technische Universit\"at Berlin, 10623 Berlin, Germany}
\author{Bella Lake}
\affiliation{Helmholtz-Zentrum Berlin f\"ur Materialien und Energie, 14109 Berlin, Germany}
\affiliation{Institut f\"ur Festk\"orperphysik, Technische Universit\"at Berlin, 10623 Berlin, Germany}
\author{Manfred Reehuis}
\affiliation{Helmholtz-Zentrum Berlin f\"ur Materialien und Energie, 14109 Berlin, Germany}
\author{A. T. M. Nazmul Islam}
\affiliation{Helmholtz-Zentrum Berlin f\"ur Materialien und Energie, 14109 Berlin, Germany}
\author{Yogesh Singh}
\affiliation{Indian Institute of Science Education and Research (IISER) Mohali, Knowledge City, Sector 81, Mohali 140306, India}
\author{Oleksandr Prokhnenko}
\affiliation{Helmholtz-Zentrum Berlin f\"ur Materialien und Energie, 14109 Berlin, Germany}
\author{Philip Pattison}
\affiliation{Swiss-Norwegian Beam Lines, European Synchrotron Radiation Facility, 71 Avenue des Martyrs, 38043 Cedex 9 Grenoble, France}
\author{S\'andor T\'oth}
\affiliation{Laboratory for Neutron Scattering, Paul Scherrer Institut, 5232 Villigen, Switzerland}

\date{\today}

\begin{abstract}
A detailed diffraction study of \cacro is presented which adds significant new insights into the structural and magnetic properties of this compound. A new crystal structure type was used where the $a$ and $b$ axes are doubled compared to previous models providing a more plausible structure where all crystallographic sites are fully occupied. The presence of two different valences of chromium was verified and the locations of the magnetic \Cr and non-magnetic \Crnm ions were identified. The \Cr ions have spin-\nicefrac{1}{2} and form distorted kagome bilayers which are stacked in an ABC arrangement along the $c$ axis. These results lay the foundation for understanding of the quantum spin liquid behavior in \cacro which has recently been reported in [C. Balz \emph{et al.}, Nature Physics, \textbf{12}, 942 (2016)].
\end{abstract}


\maketitle

\section{Introduction}

The magnetic phenomenon observable in materials arise in part from the various possible crystal structures and arrangement of magnetic ions. One important research area for which new materials are sought is topological and frustrated magnetism \cite{Cla16}. Here, structures where the magnetic ions form low-dimensional and triangular arrangements are particularly interesting as they can suppress the usual tendency for long-range magnetic order and give rise to exotic fractional excitations such as spinons or monopoles \cite{Bal10}. Triangular arrangements of magnetic ions coupled by antiferromagnetic interactions can result in frustration where it is impossible to fully satisfy all the interactions simultaneously. Two examples are triangular and kagome planes which consist of two-dimensional layers of triangles in edge-sharing and corner-sharing geometries respectively.

An example of materials which consist of triangular layers of magnetic ions are the compounds $A_{3}$Cr$_{2}$O$_{8}$ ($A$ = Sr, Ba). Here the magnetic Cr$^{5+}$ ions are coordinated by oxygen tetrahedra and have spin-$\nicefrac{1}{2}$. They form triangular bilayers in the $ab$ plane which are stacked along the $c$ axis in an ABC arrangement. In these compounds the intra-bilayer interaction which couples the two triangular layers to form the bilayer is the strongest interaction and is antiferromagnetic, as a result the spins are paired into singlets at low temperatures and the ground state is non-magnetic \cite{Sin07}. The magnetic excitations consist of breaking a singlet into a spin-1 triplet which costs a finite amount of energy, the remaining interactions allow the triplet excitations to hop from dimer to dimer giving rise to a gapped dispersive mode \cite{Kof09,Qui10}. A number of interesting physical phenomena have been observed in these materials such as field-induced transition to long-range magnetic order \cite{Kof09_2} that maps onto Bose-Einstein condensation \cite{Acz09}, and strongly correlated behavior of the magnetic excitations at elevated temperature in contrast to the thermally induced decoherence typical of convention magnets \cite{Qui12,Jen14}.

The compound \cacro which is the subject of this paper is related to the $A_{3}$Cr$_{2}$O$_{8}$ ($A$ = Sr, Ba) compounds. It also consists of Cr ions coordinated by oxygen tetrahedra, however the space group is somewhat different, $R3c$ (instead of $R\overline{3}m$) and both the $a$ and $c$ lattice parameters are doubled. As for A$_{3}$Cr$_{2}$O$_{8}$, the Cr ions are arranged on triangular bilayers which are however distorted and furthermore one in eight Cr ions is absent from the triangular bilayer. Until recently \cacro had received little attention and was mentioned in only three publications \cite{Gye81,Arc98,Gye13}. Our recent experimental investigation has however revealed that this compound has very interesting magnetic properties. It realizes a quantum spin liquid ground state \cite{Bal16} which is a topologically ordered state where the spin moments never become static but remain in collective motion down to the lowest temperatures. Spin liquids are undergoing a lot of theoretical investigation currently however very few physical realizations exist, thus \cacro provides an important addition to our current understanding of these novel states. In this paper we describe a thorough investigation of the crystal structure of \cacro including the location and valence of the magnetic Cr ions that are responsible for the magnetism and give rise to the quantum spin liquid ground state.

\begin{figure}
\centering
\includegraphics[width=0.9\columnwidth]{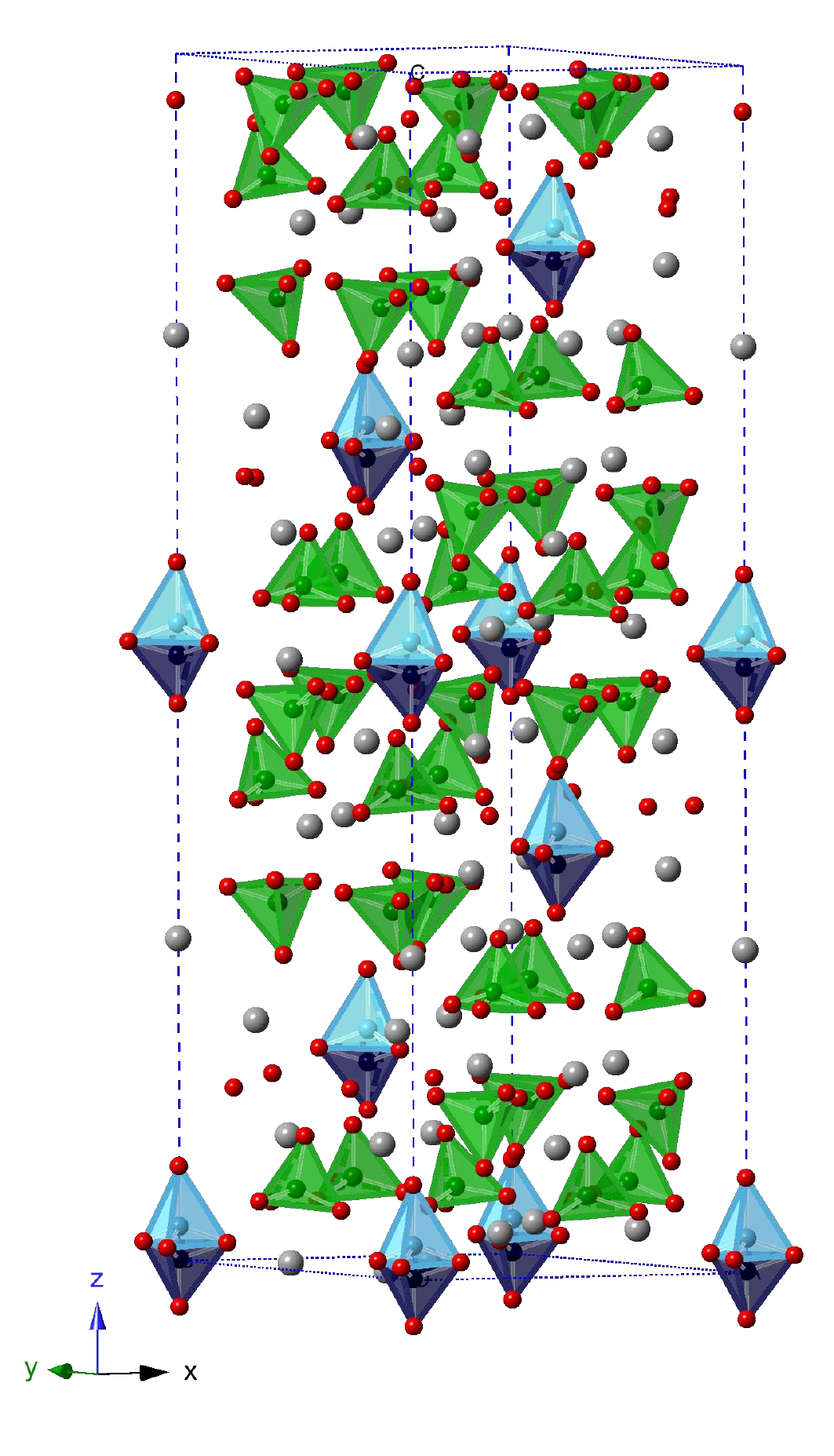}
\caption{Crystal structure of \cacro showing the conventional unit cell as given by the best model (\emph{model 3}) described in the main text. Red spheres are oxygen, gray spheres are calcium, green tetrahedra are Cr$^{5+}$O$_4$, blue and light blue tetrahedra are Cr$^{6+}$O$_4$.
\label{fig:Ca10Cr7O28_structure}}
\end{figure}

The first reported diffraction study of this compound was published in 1981 by D. Gyepesov\'{a} \emph{et al.} \cite{Gye81} where the space group is determined to be trigonal $R3c$ and the chemical composition was assumed to be Ca$_3$Cr$_2$O$_8$ by analogy to the $A_{3}$Cr$_{2}$O$_{8}$ ($A$ = Sr, Ba) compounds. Partial occupancies of one calcium and one oxygen position led to the empirical formula of Ca$_{10.07}$Cr$_7$O$_{27.58}$. This proposed crystal structure model was unsatisfactory and implied the need for further investigation. In 1998 I. Ar\v{c}on \emph{et al.} published X-ray absorption near-edge structure spectroscopy (XANES) data which suggested an average valence state of the chromium ions of 5.3(1) \cite{Arc98}. They proposed that the chromium ions in \cacro are charge ordered with two different valences, Cr$^{5+}$ and Cr$^{6+}$ in the ratio 6:1 giving an average Cr valence of 5.14 consistent with this experimental observation and they extended the chemical formula to Ca$_{10}$(Cr$^{5+}$O$_4$)$_6$(Cr$^{6+}$O$_4$). The location of the Cr$^{6+}$ could not be determined from the XANES measurement since this technique is only sensitive to the average chromium valence.

Recently in 2013 D. Gyepesov\'{a} \emph{et al.} revisited the crystal structure of this compound and published a more detailed diffraction study \cite{Gye13}. They proposed a modified structure where the partial occupancies are eliminated except at one of the CrO$_4$ tetrahedron. This tetrahedron is disordered over two possible sites whose partial occupancies sum up to full occupancy. The resulting structure is stoichiometric with chemical formula \cacro where the chromium ions have valences 5+ and 6+ in the ratio 6:1. They further suggest that it is the Cr ion in the disordered tetrahedra that has the 6+ valence.

Here we present a detailed experimental investigation of the crystal structure of \cacro to an unprecedented level of detail. We find that the 2013 model of D. Gyepesov\'{a} \emph{et al.} provides a good description of our data. However we propose a more realistic model that allows full occupancy of all sites and eliminates site disorder by introducing a supercell. Evidence for this is provided by the observation of superlattice peaks. A bond valence sum identifies the location of the Cr$^{5+}$ and Cr$^{6+}$ ions. While the Cr$^{6+}$ ions are non-magnetic, the Cr$^{5+}$ ions have spin$-\frac{1}{2}$ and magnetization and susceptibility measurements are used to confirm the ratio of these two ions in this compound. The arrangement of the magnetic Cr$^{5+}$ ions is discussed along with its consequences for the magnetic properties of \cacro.

\section{Experimental Details}

\subsection{Crystal Growth}

\begin{figure}
\centering
\includegraphics[width=\columnwidth]{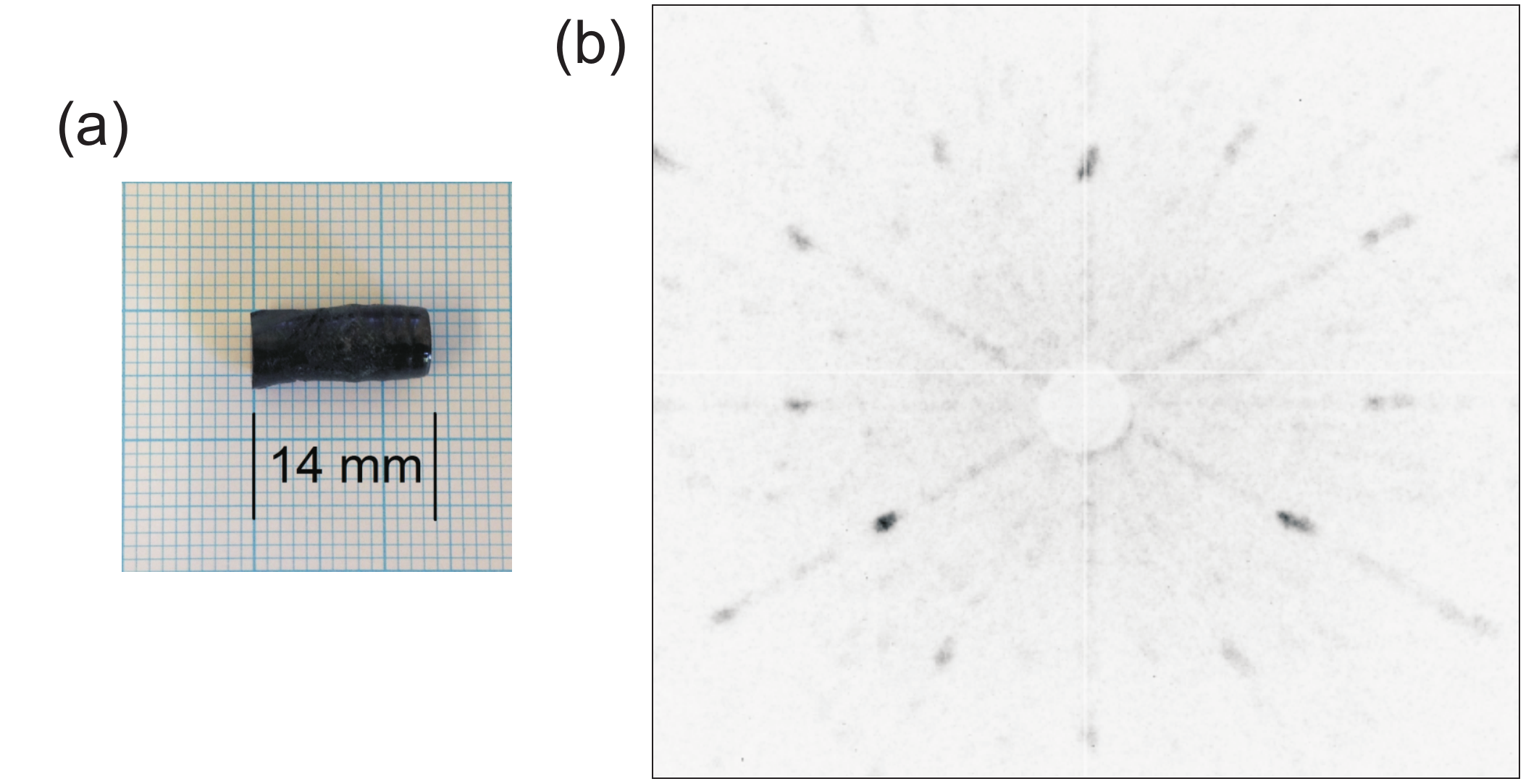}
\caption{(a) Picture of a typical Ca$_{10}$Cr$_7$O$_{28}$ single crystal with length 14~mm. (b) Neutron Laue backscattering diffraction image measured on the OrientExpress instrument at the Institut Laue-Langevin, Grenoble. The incident neutrons were parallel to the trigonal $c$ axis.\label{fig:sample_prep}}
\end{figure}

The powder samples of \cacro were prepared from high purity powder of CaCO$_3$ (99.95\%, Alfa Aesar) and Cr$_2$O$_3$ (99.97\% Alfa Aesar) following a solid state reaction route. The starting materials were mixed in the molar ratio 3:1 and then calcined in an alumina crucible in air at 1000\textdegree C for 24 hours, followed by rapid quenching to room temperature in air.
The single-crystal growth was carried out in an optical image furnace (Crystal Systems Corp., Japan) equipped with four 300~W Tungsten halide lamps focused by four ellipsoidal mirrors. To make the single crystals the obtained powder was pulverized, packed into a cylindrical rubber tube and pressed hydrostatically up to 3000~bar in a cold-isostatic-press. The resulting cylindrical rod with a diameter of 6~mm and length 7-8~cm was then sintered in air at 1010\textdegree C for 12 hours followed by rapid quenching to room temperature in air. A dense and crack-free feed rod could be obtained in this process. Since this compound is known to decompose at 959\textdegree C below the melting temperature \cite{Dev87}, the traveling-solvent-floating-zone technique was employed using an off-stoichiometric solvent to lower the melting temperature. The solvent with the composition of 28.5mol\% Cr$_2$O$_3$ - 71.5mol\% CaO was prepared using the same process as the feed rod and about 0.5~g of solvent was attached to the tip of the feed rod to start the growth. The feed rod was suspended from the upper shaft of the image furnace using nickel wire, while another smaller feed rod was fixed to the lower shaft to support the melt. Crystal growths were performed under different atmospheres of flowing air, argon, a mixture of argon and oxygen, and pure oxygen. A stable growth was achieved under an oxygen atmosphere of 0.2~MPa at a growth rate of 1~mm/h. Single crystalline rods of up to 14~mm in length and 6~mm in diameter could be obtained by this method (see figure \ref{fig:sample_prep}(a)). 

From each crystalline sample a small piece was ground and checked with X-ray powder diffraction (Bruker D8) to confirm the phase purity. Another piece of the crystal was polished and checked with a polarized light microscope to confirm the absence of any residual phase, grain boundaries or inclusions. The single crystals of Ca$_{10}$Cr$_7$O$_{28}$ were also checked with by neutron Laue diffraction. Figure \ref{fig:sample_prep}(b) shows a neutron Laue image with the trigonal $c$ axis aligned parallel to the incident beam. This picture was taken on the OrientExpress instrument at the Institut Laue-Langevin, Grenoble. Since no peak splitting of the Laue reflections is visible the single grain composition of the sample can be confirmed.

\subsection{Sample Characterization}

The crystal structure was investigated in detail using four different diffractometers. The Time-of-flight (TOF) neutron powder diffractometer EXED at the BERII reactor, Helmholtz-Zentrum Berlin (HZB) was used to collect datasets at 300\,K, 30\,K and 2\,K. Additional data at room temperature were collected on the neutron single crystal diffractometer E5 (HZB) and at the high-resolution X-ray beamline MS-Powder at the Paul-Scherrer-Institute, Villigen. The refinements were carried out using the \textsc{FullProf} software package \cite{Rod93}. Finally, high-resolution synchrotron single-crystal diffraction was carried out on the Swiss-Norwegian beam line at the European Synchrotron Radiation Facility (ESRF), Grenoble, France.

For the measurements on EXED the powder was put in a 12~mm diameter vanadium sample can. A wavelength band of $0.6<\lambda<7.8$~{\AA} was used and the detector bank was placed in backscattering geometry centered at $2\theta=151.72$\textdegree. The instrument was in high resolution mode with a Fermi chopper for primary pulse generation (FWHM $\sim$ 40 $\mu$s) \cite{Pro15}. The low temperature datasets were measured in a cryostat, and an empty cryostat measurement was used for background subtraction while the room temperature data was measured without the cryostat. In the refinement a convolution of a pseudo-Voigt with a double exponential was used and the peak shape at room temperature refined to $\sigma_1=941.032$\,$\mu$sec/{\AA}$^2$ and $\sigma_2=30.966$\,($\mu$sec/{\AA}$^2$)$^2$ for the variance of the Gaussian component and $\gamma_1=7.676$\,$\mu$sec/{\AA} and $\gamma_2=0.100$\,$\mu$sec/{\AA}$^2$ for the full-width-half-maximum (FWHM) parameters of the Lorentzian component. The back-to-back exponential decay functions yield the parameters $\alpha_0=0.063$ and $\beta_0=0.131$.

For the measurements on the 4-circle diffractometer E5, a neutron wavelength of $\lambda=0.89930$\,{\AA} was used. In order to refine the crystal structure of Ca$_{10}$Cr$_7$O$_{28}$ a full set of 3430 (1863 unique) reflections was collected at room temperature using a rod shaped single crystal with the dimensions $d=6$\,mm and $h=10$\,mm. The seed-skewness integration method was applied to determine the Bragg intensities \cite{Pet03}. The absorption correction was carried out with the program Xtal 3.4 \cite{Hal95} using a Gaussian integration with the absorption coefficient $\mu=0.32$\,cm$^{-1}$ . In the \textsc{FullProf} refinement the anisotropic Becker-Coppens extinction model was applied \cite{Bec74}. All extinction coefficients refined to very small values which indicates that extinction is negligible in Ca$_{10}$Cr$_7$O$_{28}$.

On the MS-Powder beamline, the sample was measured at room temperature in a 0.2~mm diameter glass capillary. The solid-state silicon microstrip detector MYTHEN II was used \cite{Wil13} and the capillary was mounted on a spinner for good statistical averaging. The nominal synchrotron energy of 16.0\,keV was used. The best description of the peak profile was given by split pseudo-Voigt function. In the refinement, the half-width parameters were fitted to $U_L=0.001156$, $V_L=0.000969$ for the left and $U_R=0.002344$, $V_R=0.000605$ for the right side of the peak. The ratio between Lorentzian and Gaussian contribution is $\eta=0.979$. 

In order to investigate the crystal structure of \cacro in more detail we have performed an X-ray single-crystal diffraction study on the beam line BM1A (Swiss-Norwegian) at the ESRF in Grenoble using a Pilatus2M area detector \cite{Dya16}.  For the experiments carried out at room temperature and 100~K we used a small single crystal of about 200 $\mu$m$^3$. The used wavelength of $\lambda=0.69166$~{\AA} allowed us to cover reciprocal lattice ranges $-13\leq h\leq13$, $-12\leq k\leq12$ and $-49\leq l\leq49$ where we collected a complete set of 16160 reflections.

Magnetization measurements were also performed to determine the magnetic properties of \cacro. The magnetization along the $c$ axis of up to 14 T was measured on 15~mg single crystal using a Quantum Design PPMS system at 1.8 K. DC magnetic susceptibility was measured from 400~K down to 1.8 K on a 49~mg single crystal sample using a Quantum Design MPMS Squid sensor both along the $a$ and $c$ axis. The applied DC field was 0.1~T. These experiments were performed at the Laboratory for Magnetic Measurements at HZB.

\section{Experimental results}

\subsection{Refinement in the conventional unit cell}

\begin{table*}
\caption{Comparison of the $R_F$-factors of the 4 different models for the crystal structure. The $z$ coordinates of the Cr3 and O3 ions which lie at Wyckoff position $6a(0,0,z)$ are also given.\label{tab:3_models}}
\begin{ruledtabular}
\begin{tabular}{c c c c c c c c}
& E5 & MS Powder & EXED 300\,K & EXED 30\,K & EXED 2\,K & $z$(Cr3A/Cr3B) & $z$(O3A/O3B)\\
\hline\\
original & $R_F=6.39$ & $R_F=4.75$ & $R_F=4.21$ & $R_F=3.45$ & $R_F=3.53$ & 0.00550(6) & $-0.03768(10)$ \\
model 1 & $R_F=6.44$ & $R_F=4.81$ & $R_F=4.19$ & $R_F=3.40$ & $R_F=3.40$ & 0.00565(6) & $-0.03754(5)$ \\
model 2 & $R_F=9.10$ & $R_F=6.01$ & $R_F=4.18$ & $R_F=3.80$ & $R_F=3.80$ & 0.03335(10) & 0.07759(9) \\
model 3 & $R_F=5.28$ & $R_F=3.98$ & $R_F=3.79$ & $R_F=3.14$ & $R_F=3.13$ & 0.00446(10)/ & $-0.03863(9)$/ \\
&&&&&& 0.02961(23) & 0.07192(26) \\
\end{tabular}
\end{ruledtabular}
\end{table*}

\begin{figure}
\centering
\includegraphics[width=\columnwidth]{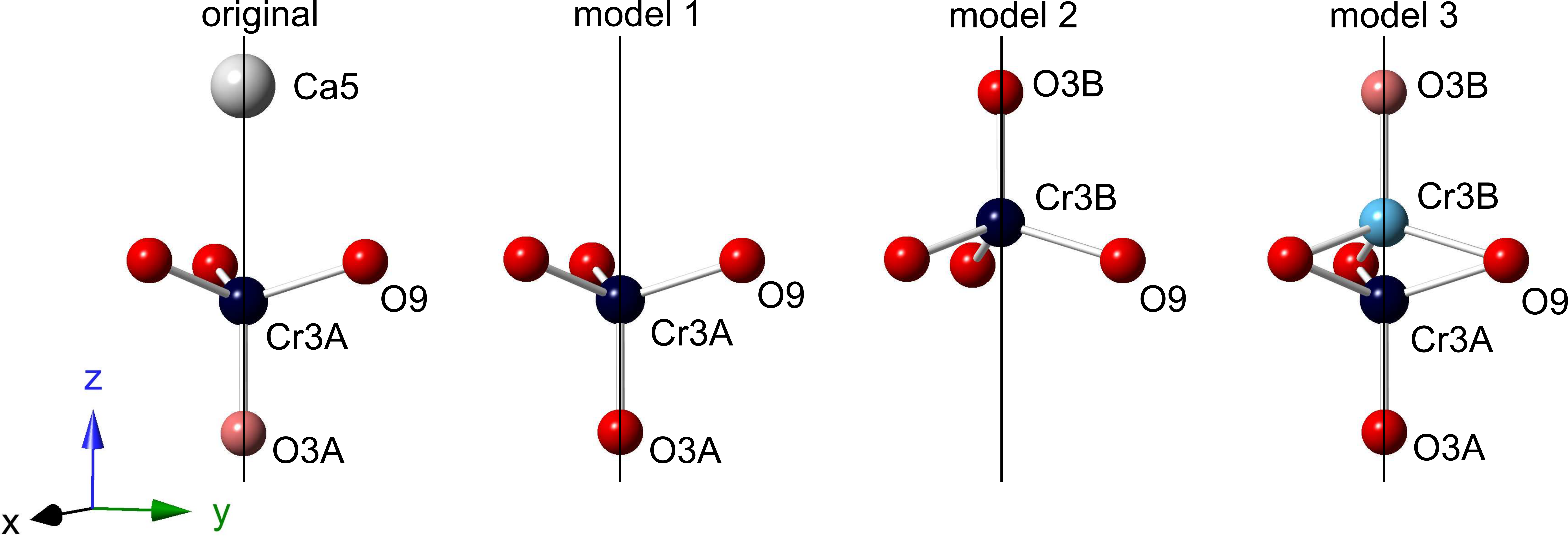}
\caption{Detail of the crystal structure around the $6a(0,0,z)$ position. In the \emph{original model} from 1981 \cite{Gye81}, Ca5 and O3A have partial occupancies of $Occ_{Ca5}=7(2)$\% and $Occ_{O3A}=58(7)$\% respectively. In \emph{model 1} and \emph{model 2}, Ca5 is eliminated and the Cr3AO$_4$ and Cr3BO$_4$ tetrahedra are fully occupied respectively. In \emph{model 3} both tetrahedra are present and their partial occupancies sum up to full occupancy. The vertical lines represent the 3-fold rotation axis parallel to $c$.\label{fig:atoms_zoom}}
\end{figure}

The structure of Ca$_{10}$Cr$_7$O$_{28}$ was refined with the conventional unit cell used for the previous refinements with lattice parameters $a=b\approx10.8$~{\AA} and $c\approx38.1$~{\AA} and trigonal space group $R3c$. Several different models were compared to the data. The first model is the \emph{original model} proposed in the 1981 paper of Gyepesov\'{a} \emph{et al.} \cite{Gye81} who started with the chemical formula Ca$_{3}$Cr$_{2}$O$_{8}$. This model is unsatisfactory because a number of the atomic sites refined to partial occupancies particularly in the vicinity of the Wyckoff position $6a(0,0,z)$ located on the 3-fold axis parallel to the $c$ axis. Here the occupancy of Ca5 is $Occ_{Ca5}=7(2)$\%\ and the occupancy of O3A is $Occ_{O3A}=58(7)$\%\ (see Fig.~\ref{fig:atoms_zoom}a) giving chemical formula Ca$_{10.07}$Cr$_7$O$_{27.58}$. 

\begin{figure}
\centering
\raisebox{2 mm}{\includegraphics[width=\columnwidth]{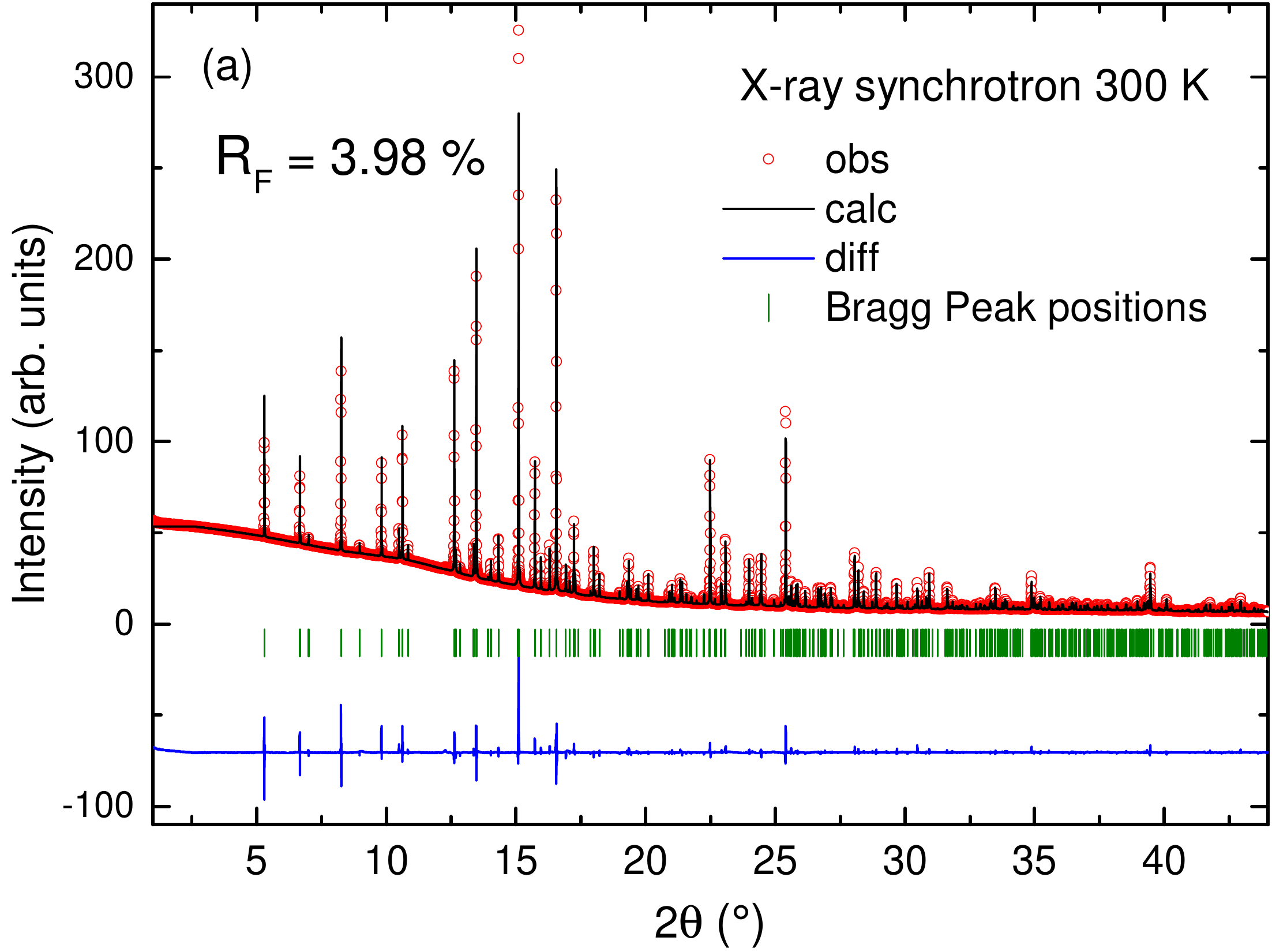}}\quad
\includegraphics[width=0.8\columnwidth]{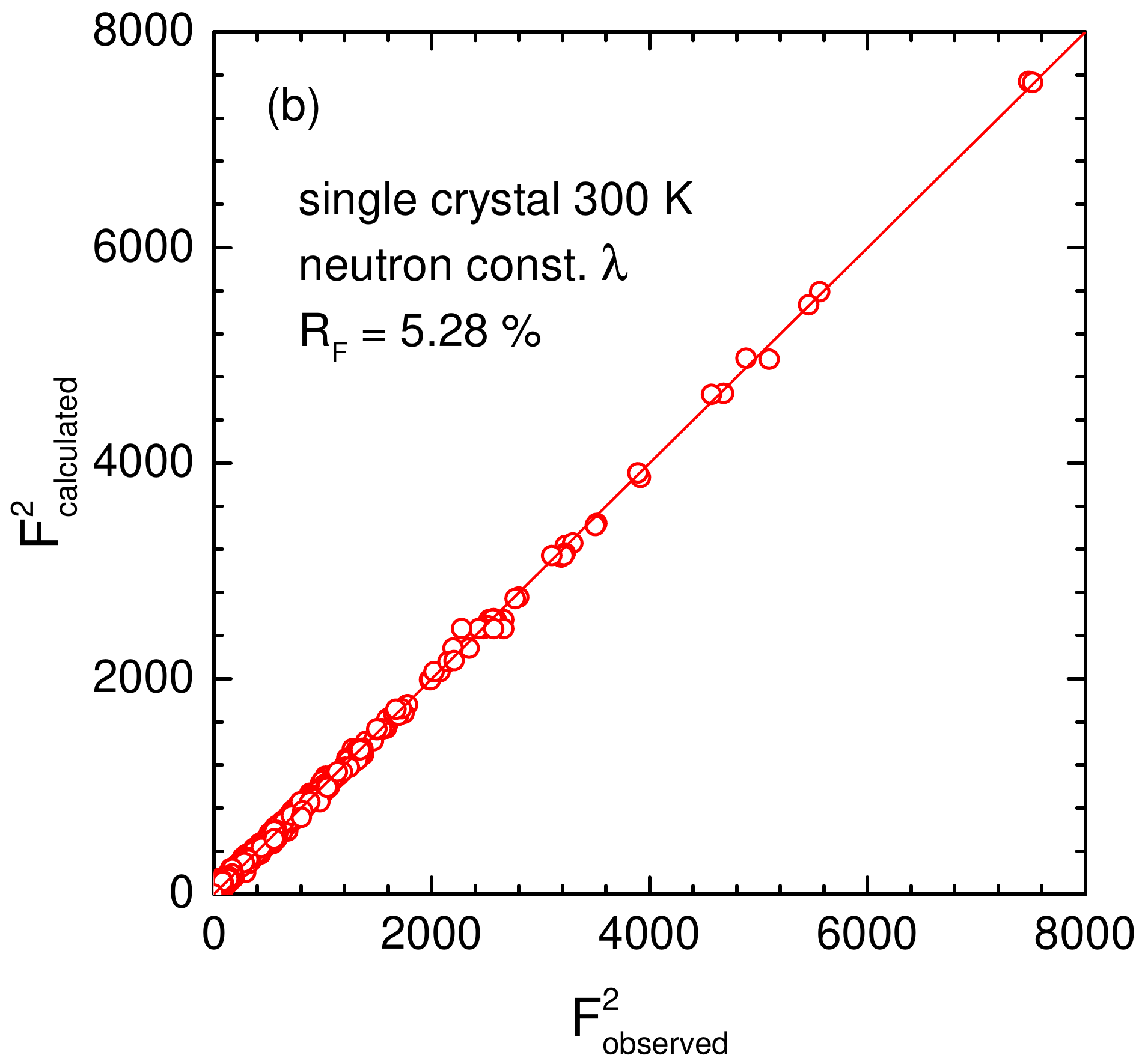}
\caption{(a) The observed (red circles) and calculated (black lines) X-ray synchrotron powder diffraction pattern for Ca$_{10}$Cr$_7$O$_{28}$ using the best model (\emph{model 3}) described in the text. The difference between observed and calculated intensities is shown by the blue line at the bottom of the panel. The green vertical bars show the Bragg peak positions. (b) Comparison between the observed and calculated structural peak amplitudes for the E5 single crystal neutron refinement using \emph{model 3}.}\label{fig:FullProf}
\end{figure}

Starting from this model but fixing the almost unoccupied Ca5 to $Occ_{Ca5}=0$\% and the former approximately half occupied O3A to $Occ_{O3A}=100$\% to yield a fully occupied Cr3AO$_4$ tetrahedron gives \emph{model 1} (Fig.~\ref{fig:atoms_zoom}b). The advantage of this model is that it avoids partial occupancies and is now stoichiometric giving the chemical formula Ca$_{10}$Cr$_7$O$_{28}$. However, differential fourier analysis using the single crystal data to calculate the scattering density inside the unit cell reveals a rest density around the position $(0,0,0.05)$ which is not accounted for by \emph{model 1}. This is approximately at the position where Ca5 was located in the \emph{original model}. To solve this problem, the rest density was ascribed to two new crystallographic positions labeled Cr3B and O3B. They form a new Cr3BO$_4$ tetrahedron which shares the three symmetry-related O9 atoms in the $ab$ plane with the Cr3A tetrahedron. In \emph{model 2}, which is the inverse of \emph{model 1}, the new positions Cr3B and O3B are fully occupied while Cr3A and O3A are empty (Fig.~\ref{fig:atoms_zoom}c). Finally, in \emph{model 3} the CrO$_4$ tetrahedron can occupy both positions. The occupancies of the Cr3AO$_4$ and Cr3BO$_4$ tetrahedra are constrained so that the sum of their occupancies is equal to 100\%\ (Fig.~\ref{fig:atoms_zoom}d). This model was also used to refine the structure in Ref.\ \cite{Gye13}.

These 4 models were refined using all the measurements individually. The goodness of fit (the $R_F$-factors) for the 4 models are given in Table \ref{tab:3_models} and reveal that \emph{model 3} agrees best with all datasets. This confirms that the true stoichiometry is indeed Ca$_{10}$Cr$_7$O$_{28}$ and that at least in the conventional unit cell the model including inversion twinning yields the best refinement. The trigonal space group $R3c$ reported in literature \cite{Gye81,Gye13} could also be confirmed. 

To obtain the most accurate room temperature structure of Ca$_{10}$Cr$_7$O$_{28}$, the lattice parameters were refined from the synchrotron X-ray powder data and fixed in the neutron single crystal refinements while the atomic positions were refined from the neutron single crystal data and then likewise fixed in the X-ray powder refinements. The thermal parameters were refined for both datasets individually. These datasets along with the best fits are shown in figure~\ref{fig:FullProf}.

The refined atomic positions, thermal parameters and lattice parameters for \emph{model 3} at room temperature are presented in table \ref{tab:refinement} (see appendix). The crystal structure of Ca$_{10}$Cr$_7$O$_{28}$ is shown in figure \ref{fig:Ca10Cr7O28_structure}. Since the Cr3BO$_{4}$ tetrahedron (light blue) is an inversion twin of the Cr3AO$_{4}$ tetrahedron (dark blue) only one of them is occupied at a given site and their ratio refined to Cr3A/Cr3B\,=\,0.726(12)/0.276(12) from the single crystal data. The same ratio applies to O3A/O3B. In the final refinement the occupancies of all other positions were fixed to 100\,\%. Their full occupancy was confirmed beforehand by allowing partial occupancies which all refined to full occupancy within error bar.

\begin{figure}
\centering
    \includegraphics[width=\columnwidth]{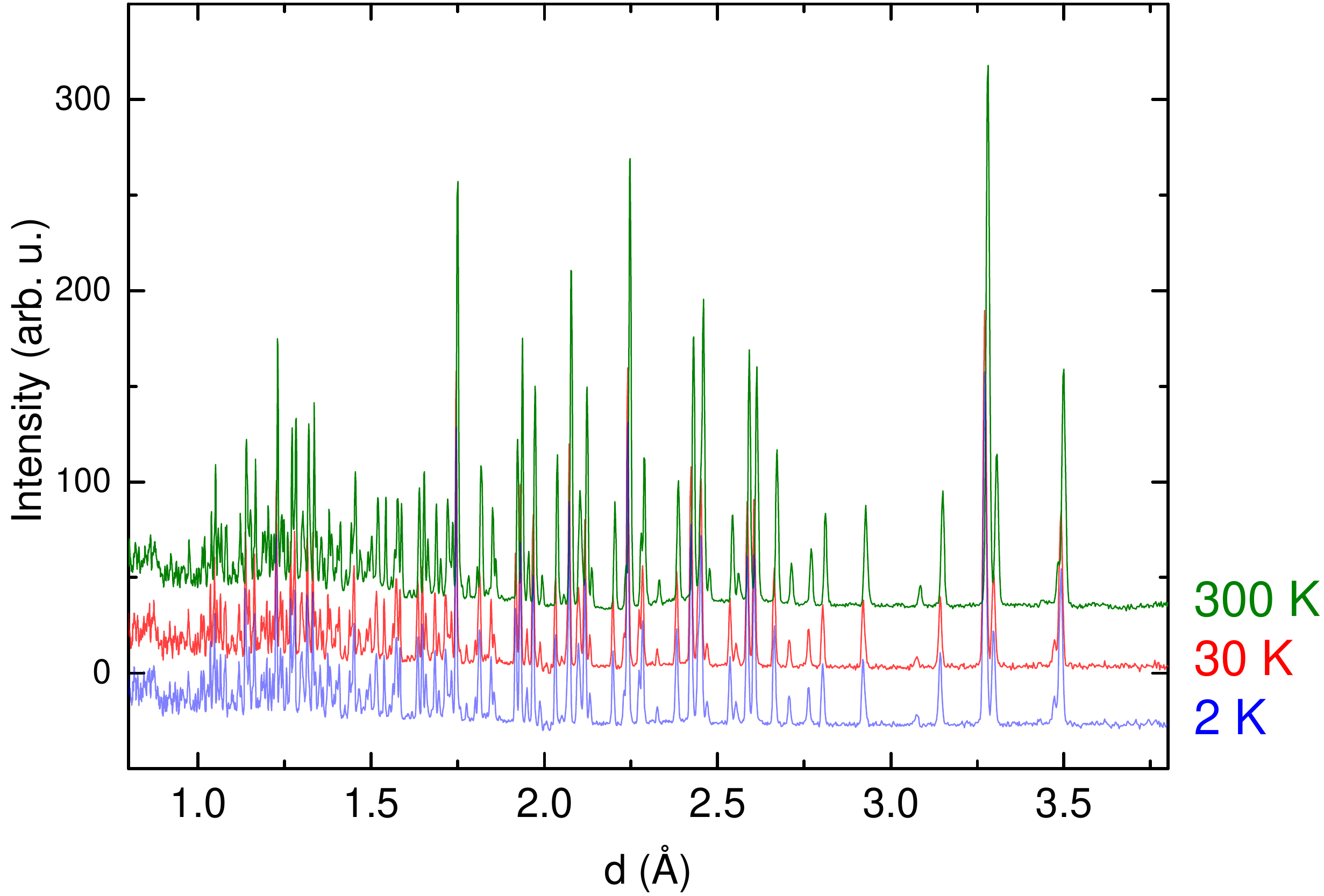}
\caption{Observed neutron TOF powder diffraction patterns for Ca$_{10}$Cr$_7$O$_{28}$ at 2 K, 30 K and 300 K. The patterns have been shifted vertically for comparison. No additional peaks or peak broadening appear between room temperature and 2\,K. The 300~K data has an inherently higher signal to background ratio since no cryostat was used for this measurement.}\label{fig:V15_comp}
\end{figure}

\begin{figure}
\centering
\includegraphics[width=\columnwidth]{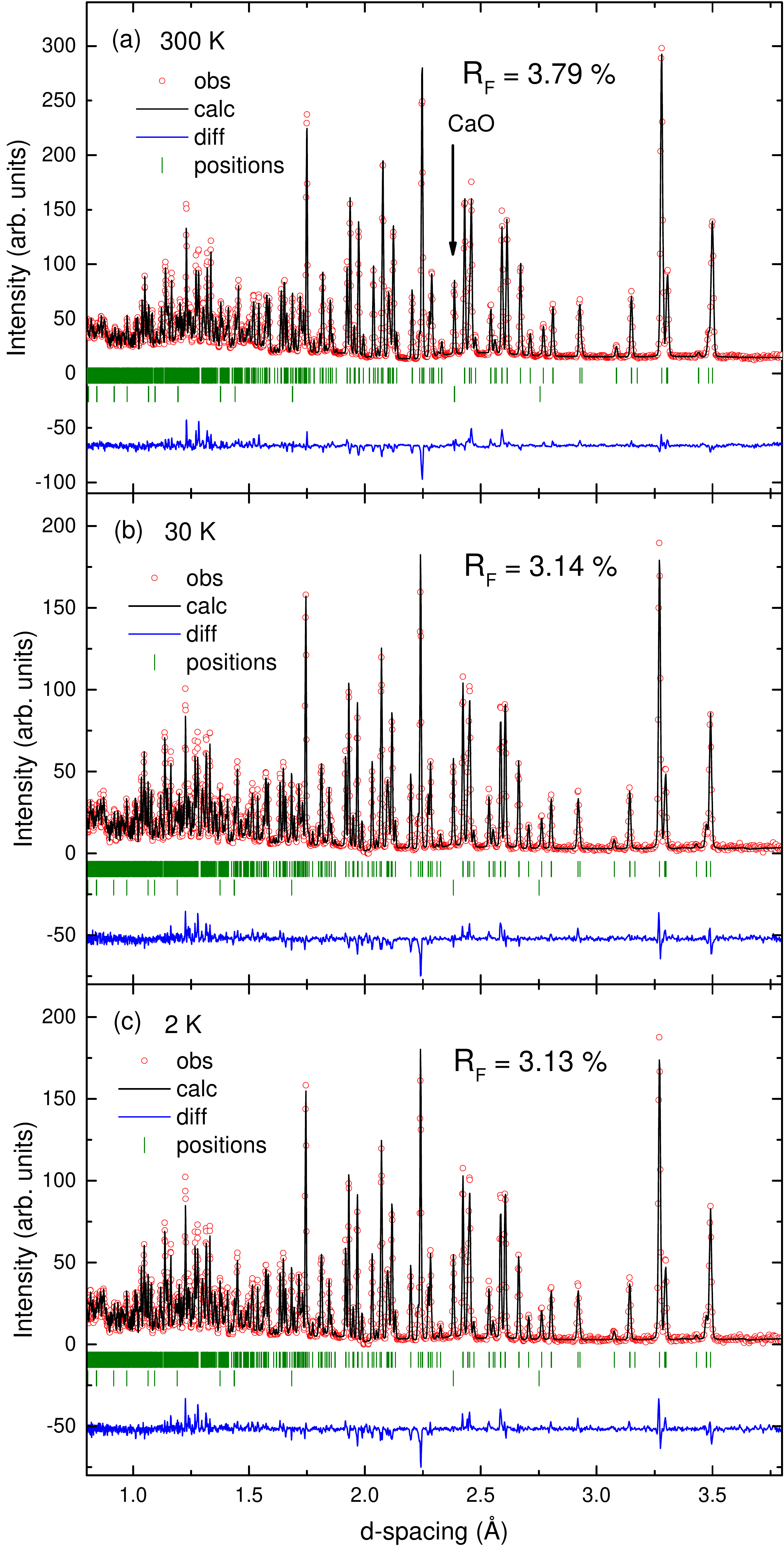}
\caption{The observed (red circles) and calculated using \emph{model 3} (black lines) neutron TOF diffraction pattern for Ca$_{10}$Cr$_7$O$_{28}$ at (a) 300~K, (b) 30~K and (c) 2~K. The difference between observed and calculated intensities is shown by the blue line at the bottom of the panel. The green vertical bars show the Bragg peak positions. The second row of the peak markers shows the CaO impurity phase which has an intense peak at 2.4~{\AA}.}\label{fig:FullProf2}
\end{figure}

The neutron TOF powder diffraction patterns recorded at 300\,K, 30\,K and 2\,K were used to check for any phase transition or lattice distortion below room temperature. Overplotting the observed patterns in figure \ref{fig:V15_comp} shows that no new peaks, peak splitting or significant changes in intensity occur. This implies the absence of any structural distortion in Ca$_{10}$Cr$_7$O$_{28}$ down to 2 K. These datasets were refined using \emph{model 3} and the best fits are shown along with the data in figure~\ref{fig:FullProf2}. An earlier batch of powder was used for the neutron TOF measurement than that used for the X-ray synchrotron measurement, which contained a CaO impurity phase. A strong peak caused by this impurity is indicated by the black arrow in figure~\ref{fig:FullProf2}(a). The weight percentage of the impurity phase refined to 13\%.

\begin{table}
\caption{Refinement of the lattice parameters from the EXED neutron TOF powder data at 3 different temperatures using \emph{model 3}.\label{tab:V15_ref}}
\begin{ruledtabular}
\begin{tabular}{c c c}
& $a$ & $c$ \\
300~K & 10.68873(239) & 37.80577(973) \\
30~K & 10.65156(135) & 37.70086(561) \\
2~K & 10.65167(137) & 37.70106(568) \\
\end{tabular}
\end{ruledtabular}
\end{table}

The refined lattice parameters at the three temperatures are given in table~\ref{tab:V15_ref}. The room temperature lattice parameters refined from the EXED data are systematically smaller by about 1\% compared to those derived from the synchrotron data (cf. table \ref{tab:refinement}). This is an experimental effect due a very subtle dependence of the lattice parameters on the instrument calibration in a time-of-flight experiment which has been observed previously (see e.g. Ref.~\cite{Pro13}). Besides the uncertainty in the absolute values, the EXED refinement gives precise information about the relative change of the lattice parameters between 2~K and 300~K which is found to be less than 0.5\% . Furthermore, the fractional coordinates are constant within error bar as a function of temperature.

\subsection{Refinement in the supergroup}

Now we discuss the refinement of our diffraction data in an enlarged unit cell. As discussed in the previous section, the CrO$_4$ group on the threefold axis in the vicinity of $6a(0,0,z)$ is disordered over two possible orientations in the conventional unit cell. The refined occupancies of Cr3A and O3A were 0.695(6), while those of Cr3B and O3B were 0.305(6) in the paper of  Gyepesov\'{a} \emph{et al.} \cite{Gye13}. In our single-crystal neutron diffraction study presented here, the refinements showed a significantly higher occupancy of 0.726(12) for Cr3A and O3A, and a correspondingly lower occupancy of 0.276(12) for Cr3B and O3B. Interestingly, these values are very close to the ideal values of \nicefrac{3}{4} and \nicefrac{1}{4}.

\begin{figure}
\includegraphics[width=\columnwidth]{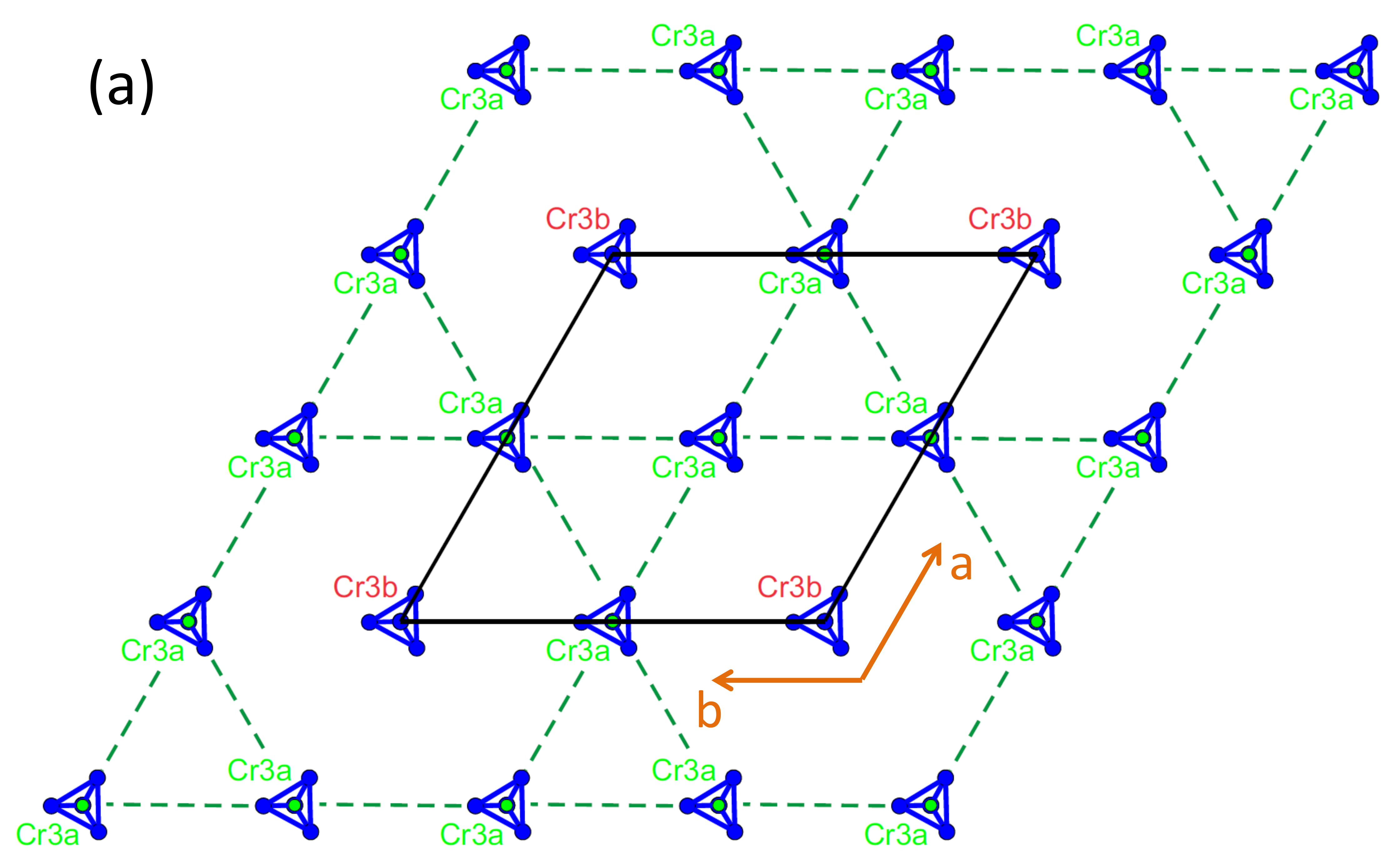}\\
\vspace{2 mm}
\includegraphics[width=\columnwidth]{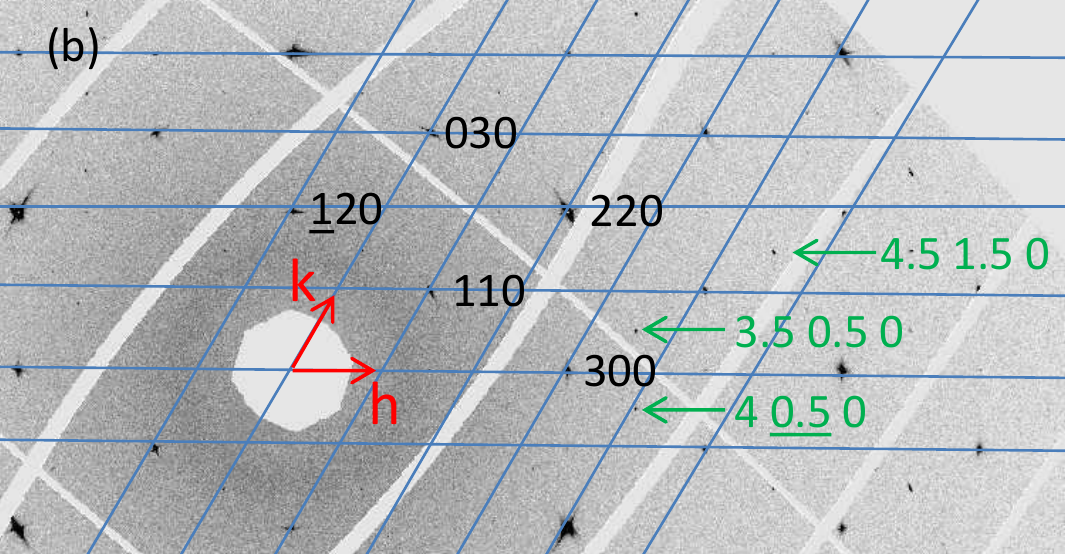}\\
\vspace{2 mm}
\includegraphics[width=\columnwidth]{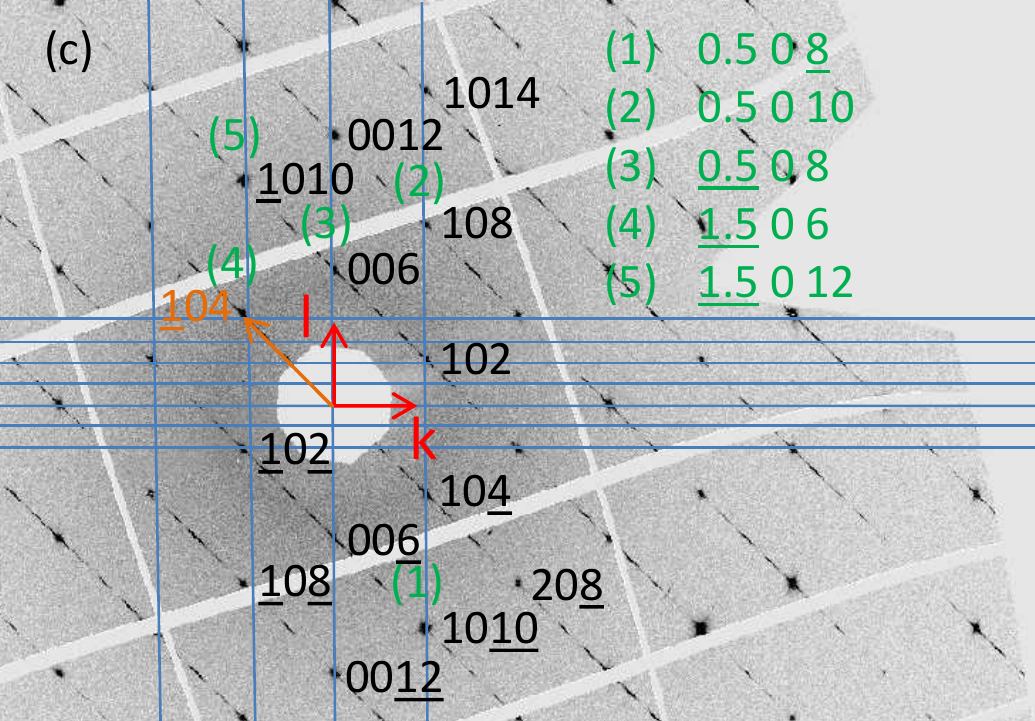}
\caption{(a) Schematic of the real space supercell (solid black lines) of \cacro in the hexagonal $ab$ plan close to $c=0$. Compared to the conventional unit cell, the $a$ and $b$ axes are doubled and the Cr3A and Cr3B positions are ordered in a 3:1 ratio. (b) The diffraction pattern in the reciprocal $hk0$ plane obtained from single crystal X-ray diffraction at room temperature. The grid shows the Brillouin zone according to the conventional unit cell and characteristic structural reflections are labeled. The green arrows show examples of observed reflections with half integer values of $h$ and/or $k$. (c) The diffraction pattern in the reciprocal $h0l$ plane. The reflections with half integer coordinates are smeared out in the $[\bar{1}04]$ direction (orange arrow) and almost appear as lines of scattering. \label{fig:superstructure}}
\end{figure}

Assuming that the Cr3A and Cr3B tetrahedra have the ratio of 3:1, it is possible to define a crystal structure with an enlarged unit cell where all the Cr3 and O3 sites are fully occupied by doubling the $a$ and $b$ lattice parameters. Group-subgroup relations show that this superstructure can also be described in the same trigonal space group $R3c$, where the new basis vectors of the real space lattice are $a'=-2a$, $b'=-2b$, and $c'=c$. In the new setting the $h'k'l'$ values of a reflection are changed to $(-2h,-2k,l)$ in relation to the old ones. The Cr3B and O3B ions which originally had occupancy \nicefrac{1}{4} can be placed at the Wyckoff position $6a(0,0,z')$ where $z'$=$z$ reaching now the full occupancy. Accordingly, the Cr3A and O3A atoms which had occupancy \nicefrac{3}{4} can be placed at the position $18b(x',y',z')$ where the positional parameters are $x'$=$y'$=\nicefrac{1}{4}, $z'$=$z$. This new representation finally leads to full occupancies at all sites and a schematic of the $ab$ plane is shown in figure~\ref{fig:superstructure}(a). In the case of the Cr3A atoms (green) the apical oxygen atoms are located below the hexagonal plane, while in the case of the other chromium site the apical oxygen are located above hiding in this plot the Cr3B atoms. The in-plane oxygen positions are identical for Cr3A and Cr3B. The Ca4 atoms which also occupied the position $6a(0,0,z)$ in the old setting, split into the two positions $6a(0,0,z')$ and $18b(x',y',z')$ with $x'$=$y'$=\nicefrac{1}{4}, $z'$=$z$ and are now labeled as Ca4B and Ca4A respectively. The $x'$ and $y'$ parameters of the atoms Ca4A, Cr3A, and O3A were not allowed to vary during the refinement in the new setting. For the Cr3 and O3 atoms, the constraints $z$(Cr3A)~=~$-$~$z$(Cr3B) and $z$(O3A)~=~$-$~$z$(O3B) were used. The other atoms Ca1, Ca2, Ca3, Cr1, Cr2, and O1 - O9 which are located at $18b(x,y,z)$ in the old setting now split into four sites all at $18b(x',y',z')$. In the new setting these atoms are labeled as Ca1A, Ca1B, Ca1C, Ca1D, etc. to distinguish between them in the refinement. The coordinates of their atomic positions are transformed as follows: $x'=(1-x)/2$, $y'=(1-y)/2$, and $z'=z$. Due to the fact that the number of parameters is considerably enlarged the following constraints were used for the atoms located at $18b(x',y',z')$: Atoms labeled with A get the coordinates $(x',y',z')$, atoms B $(x'+\frac{1}{2},y',z')$, atoms C $(x',y'+\frac{1}{2},z')$, and atoms D $(x'+\frac{1}{2},y'+\frac{1}{2},z')$. Further the anisotropic thermal parameters of each set (A,B,C,D) were constrained to be equal during the refinements.

These constraints are justified because they ensure that the positions of all the ions except the $z$-coordinate of Cr3 and O3 are identical within each of the 4 original unit cells that make one supercell. In the end the number of refineable parameters could be reduced to the same number as for the refinement of \emph{model 3} and the refinement of this supergroup structure consequently results in the same residuals as found for \emph{model 3}. Thus it is not possible to distinguish between these models simply by the goodness of fit. Nevertheless the supergroup provides a more physical description of Ca$_{10}$Cr$_7$O$_{28}$ as it allows full occupancy of all sites.

One consequence of the supercell is that superlattice peaks at ($\frac{1}{2}$,$\frac{1}{2}$,0) positions (in the coordinate system of the original cell) are expected due to the Cr3A/Cr3B and O3A/O3B ordering which doubles the unit cell along both the $a$ and $b$ directions. These superlattice peaks are not observable in the powder measurements or in the neutron single crystal data probably because they have very weak intensity. In order to search for these superlattice peaks we have performed high resolution/ high intensity X-ray synchrotron single-crystal diffraction at the Swiss-Norwegian beam line at the ESRF, Grenoble, France. Indeed very weak additional reflections corresponding to a lattice doubling of $a$ and $b$ were observed confirming the correctness of the superstructure model. The reciprocal $hk0$ scattering plane is shown in figure~\ref{fig:superstructure}(b). Super-structure reflections are visible which obey the reflection condition $-h'+k'+l=3n$ of the space group of the enlarged unit cell (also $R3c$), where $h'$ and $k'$ are the reciprocal lattice units of the supercell and can be written in terms of the reciprocal lattice units $h$ and $k$ of the conventional cell as $h'=2h$, $k'=2k$. The new reflections are also visible in the projection onto the $h0l$ plane shown in figure~\ref{fig:superstructure}(c). In this plane it becomes obvious that the superstructure reflections have a diffuse character almost forming lines of scattering along the [$\bar{1}04$] direction.

\begin{table*}
\caption{Cr-O bond distances $d$ of the four structurally different CrO$_4$ tetrahedra and Bond Valence Sum calculation for the Cr ions. The spin value of each ion is also given.\label{tab:BVS}}
\begin{ruledtabular}
\begin{tabular}{l l l l l l l l}
\multicolumn{3}{l}{Interatomic distances [{\AA}]}\\
\hline
$d$(Cr1-O1)	& 1.666(3)&	$d$(Cr2-O5)	& 1.690(3)	&	$d$(Cr3A-O3A)	& 1.653(6)	&	$d$(Cr3B-O3B) &	1.611(19) \\
$d$(Cr1-O2) &	1.677(4) & $d$(Cr2-O6)	& 1.716(6)	&	$d$(Cr3A-O9)	& 1.650(7)	&	$d$(Cr3B-O9)	& 1.656(8) \\
$d$(Cr1-O3) &	1.733(3)	&	$d$(Cr2-O7)	& 1.701(2)	&	$d$(Cr3A-O9)	& 1.650(7)	&	$d$(Cr3B-O9)	& 1.656(8) \\
$d$(Cr1-O4)	& 1.722(5)	&	$d$(Cr2-O8)	& 1.674(6)	&	$d$(Cr3A-O9)	& 1.650(7)	&	$d$(Cr3B-O9)	& 1.656(8)\\
\hline\hline
Ion & BVS & nominal & spin & Ion & BVS & nominal & spin\\
\hline
Cr1 & 4.85+ & 5+ & \nicefrac{1}{2} & Cr3A & 5.89+ & 6+ & 0\\
Cr2 & 4.90+ & 5+ & \nicefrac{1}{2} & Cr3B & 6.00+ & 6+ & 0\\
\end{tabular}
\end{ruledtabular}
\end{table*}

The observation of elongated spots parallel to [$\bar{1}04$] instead of sharp Bragg peaks at half integer positions indicates some amount of remaining disorder among the Cr3A and Cr3B positions in the plane perpendicular to the $[\bar{1}04]$ direction even in the enlarged unit cell. Since sharp half-order Bragg peaks have been observed in the $hk0$ plane, the superstructure is well ordered in the $ab$ plane as shown in figure~\ref{fig:superstructure}(a). Therefore the diffuse character of the peaks must be caused by stacking faults along the $c$ axis. Unfortunately, the weakness of the superstructure reflections and their diffuse character did not allow us to use them in a crystal structure refinement. The fact that both reciprocal planes presented in figure~\ref{fig:superstructure} show additional weak reflections only at the positions which fulfill the reflection condition $-h'+k'+l=3n$ of the supercell however confirms the proposed enlarged unit cell. Lastly it should be noted that the superstructure reflections as well as the diffuse scattering remain unchanged down to the lowest measured temperature of 100~K.

\subsection{Bond Valence Sum}

The chemical formula of Ca$_{10}$Cr$_7$O$_{28}$ suggests that the chromium ions must have an average valence of +5.143 in order to balance the +2 valence of the Ca$^{2+}$ ions and the -2 valence of the O$^{2-}$ ions. As an insulating compound we propose that the Cr ions are charge ordered with a ratio of Cr$^{5+}$ to Cr$^{6+}$ of 6:1. It is clearly interesting to know which of the Cr ions have valence +5 and which have valence +6. This can be achieved by calculating the Bond Valence Sum (BVS) using our refined interatomic distances. The Cr-O bond distances within each CrO$_{4}$ tetrahedra are listed in table \ref{tab:BVS}. The calculation of the BVS for the four inequivalent chromium ions Cr1, Cr2, Cr3A and Cr3B, was performed using these Cr-O distances and the formula
\begin{align}
\text{BVS}=\sum_i\text{e}^{(d_0-d_i)/B}
\end{align}
where $d_0$ is the ideal bond length, $d_i$ the observed bond length and $B=0.37$, an empirical parameter for chromium ions. The values $d_0=1.77$~{\AA} and $d_0=1.794$~{\AA} for Cr$^{5+}$ of and Cr$^{6+}$ respectively were taken from I. D. Brown \emph{et al.} \cite{Bro85}. The results of the BVS calculation are also shown in table \ref{tab:BVS}. The BVS values of Cr3A and Cr3B are close to +6 while those of Cr1 and Cr2 are close to +5. Since the number ratio of the different chromium ions is Cr1:Cr2:Cr3 = 3:3:1, the Cr1 and Cr2 sites together account for six Cr$^{5+}$ ions while the Cr3A/Cr3B sites together account for only one Cr$^{6+}$ ion, therefore the average valence from the BVS calculation is $(5*6+1*6)/7=36/7=5.143$. The chemical formula of \cacro can hence also be written as Ca$_{10}$(Cr$^V$O$_4$)$_6$(Cr$^{VI}$O$_4$). This result confirms the charge ordering suggested in Ref.\ \cite{Gye13} and is also consistent with the results of a previous XANES measurements which found an average Cr valence of 5.3(1) \cite{Arc98}.
	
\subsection{Magnetic Measurements}

\begin{figure}
\includegraphics[width=\columnwidth]{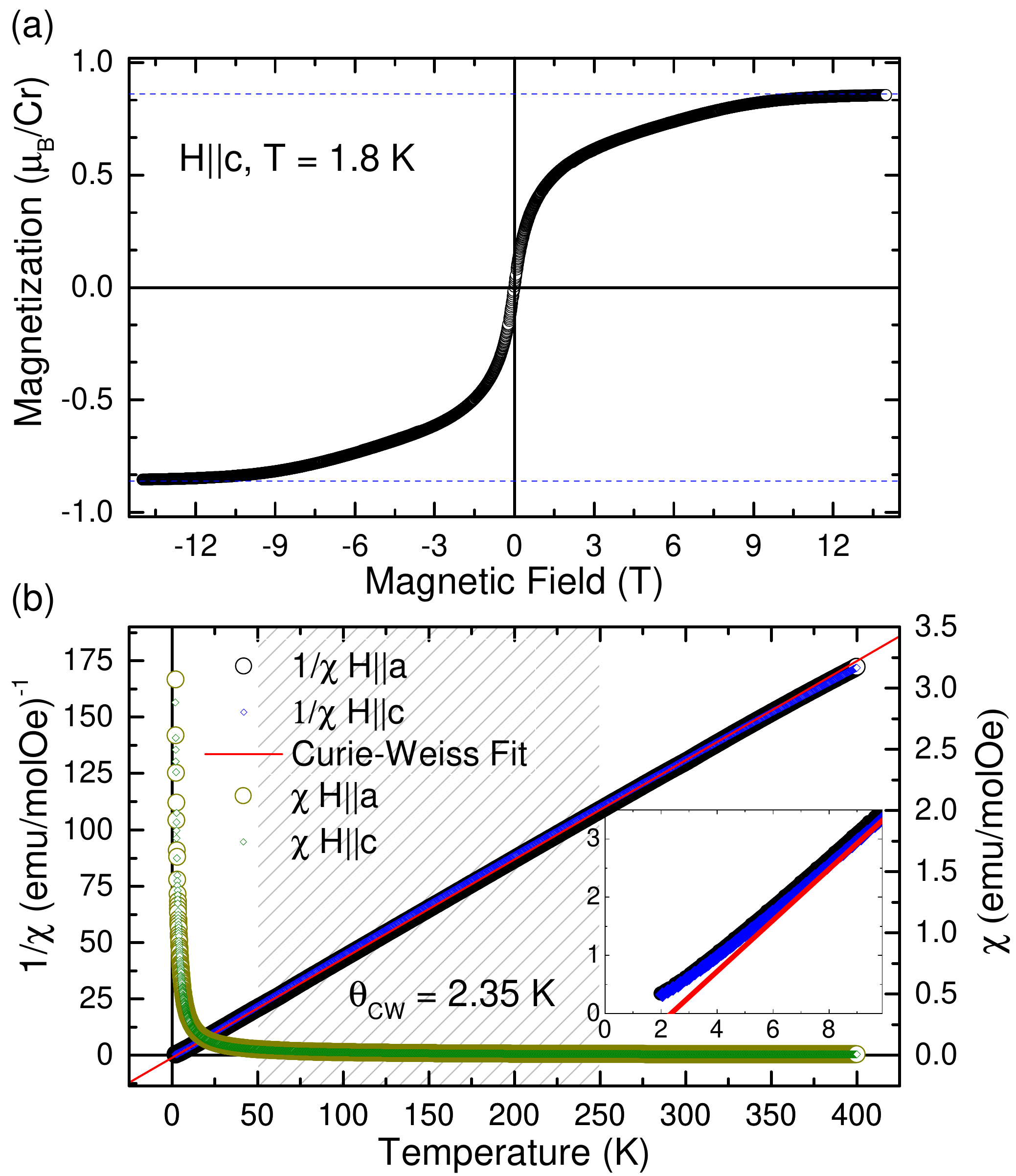}
\caption{(a) Magnetization per Cr ion for the field applied along the $c$ axis at 1.8~K. (b)Susceptibility (right axis) and inverse susceptibility (left axis) measured from 1.8 - 400~K for a field of 0.1~T parallel and perpendicular to the $c$ axis. The Curie-Weiss fit of $1/\chi$ yields $\theta_{CW}=2.35$~K as shown in the inset.\label{fig:mag_sus}}
\end{figure}

The chromium charge ordering has important implications for the magnetic properties of Ca$_{10}$Cr$_7$O$_{28}$. Cr3A and Cr3B with oxidation state +6 have no electrons in their $3d$-shell while Cr1 and Cr2 with oxidation state +5 have one unpaired electron ($3d^{1}$). Thus Cr3A and Cr3B have spin zero and are non-magnetic whereas Cr1 and Cr2 have spin-$\nicefrac{1}{2}$. To learn more about the magnetic properties we performed magnetization and DC susceptibility measurements. Figure~\ref{fig:mag_sus}(a) shows the magnetization at 1.8~K with field applied along the $c$ direction. The magnetization increases rapidly and then becomes constant above $\approx 12$ T showing that this field is strong enough to overcome the interactions between the magnetic Cr$^{5+}$ ions and force them to point along the field direction. Since the Cr$^{5+}$ ions have spin $S=\nicefrac{1}{2}$ they will each contribute $g_s \mu_B S = 1 \mu_B$ (assuming $g_s = 2$) to the magnetization, while the Cr$^{6+}$ ions with spin $S=0$ will not contribute. Since the ratio of Cr$^{5+}$ to Cr$^{6+}$ is 6:1, the expected saturation magnetization per Cr ion is $6/7 \mu_{B} = 0.857\mu_{B}$. The experimental value of the saturation magnetization 0.851$\mu_{B}$ is in good agreement with the expected one confirming the predicted spin values and valences of the Cr ions. 

Further evidence for these spin values and valences comes from DC susceptibility measurements. Figure~\ref{fig:mag_sus}(b) shows the susceptibility $\chi$ and inverse susceptibility $1/\chi$ measured with a field of 0.1~T. The inverse susceptibility follows a straight line suggesting Curie-Weiss behavior. The data in the temperature range 50-250 K was fitted by the Curie-Weiss law $\chi = C / (T-T_{CW})$, giving the Curie-Weiss temperature $T_{CW} = 2.35$~K and the Curie constant $C=0.38$. The Curie constant can be used to extract the effective moment ($\mu_{eff}=\sqrt{8C}\mu_{B}$) and gives $\mu_{eff} = 1.74\mu_{B}$ per \Cr ion assuming 6 out of every 7 Cr ions have $S=\nicefrac{1}{2}$ while the 7$^{th}$ has $S=0$. The expected value of $\mu_{eff}$ for a spin-\nicefrac{1}{2} ion is $\mu_{eff} = \frac{6}{7} g_s \sqrt{S(S+1)} \mu_{B} = 1.73 \mu_{B}$ again in good agreement with the experimental result. 

Finally it should be mentioned that the susceptibility is independent of field direction. This indicates the absence of significant magnetic anisotropy in this compound which in turn suggest that the orbital moment of the Cr$^{5+}$ ions in Ca$_{10}$Cr$_7$O$_{28}$ is almost fully quenched. While substantial quenching of orbital moment in transition metal ions such as chromium is quite common due to their strong crystal field compared to their spin-orbit coupling, there is often a weak residual anisotropy due to a small amount of unquenched moment. As mentioned before, Cr$^{5+}$ ions have a single unpaired electron in the $3d$-shell, because of the tetrahedral crystal field produced by the surrounding O$^{2-}$ ions this electron will occupy the lower lying $e_{g}$ orbitals. Furthermore because of the highly distorted nature of the CrO$_{4}$ tetrahedron for all Cr sites we expect that the $e_{g}$ doublet is split into two non-degenerate levels with the electron occupying the lowest level (see Table \ref{tab:BVS} which shows that the Cr1-O and Cr2-O bond distances are all different). Absence of orbital degeneracy for a single electron in the $3d$ shell is expected to lead to a highly quenched orbital moment and a very isotropic magnetic moment with spin$-\frac{1}{2}$ as observed experimentally.

\section{Conclusion}

\begin{figure}
\includegraphics[width=\columnwidth]{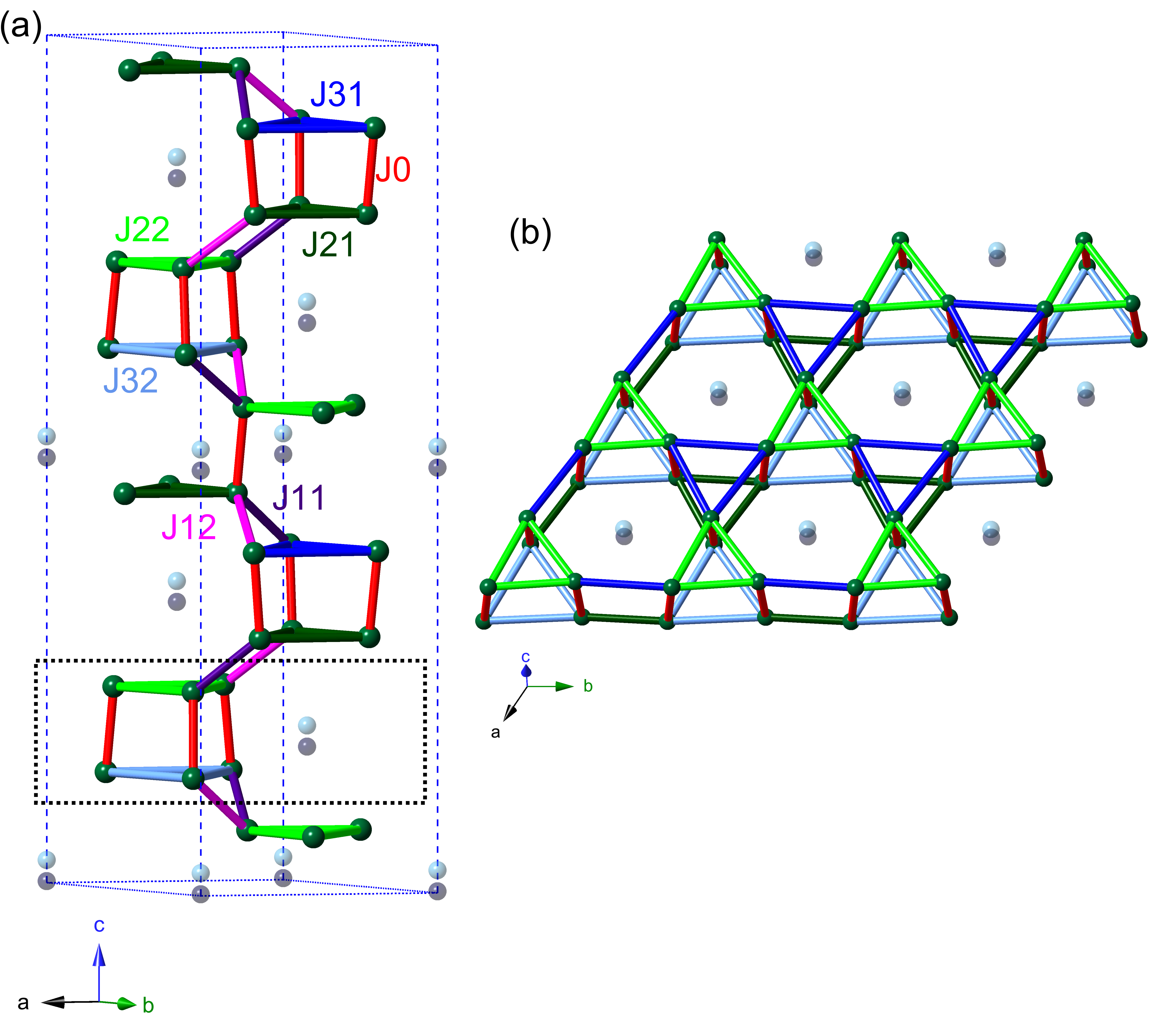}
\caption{(a) Conventional unit cell showing only the chromium ions. The color code for Cr1, Cr2, Cr3A and Cr3B is the same as in Fig.~\ref{fig:Ca10Cr7O28_structure}: Cr1 and Cr2 are green, Cr3A and Cr3B are blue and light blue. Cr1 and Cr2 are magnetic with spin \nicefrac{1}{2} while Cr3A and Cr3B are non-magnetic. The seven shortest magnetic exchange paths $J_0-J_{32}$ between Cr1 and Cr2 are indicated by the colored bonds. The dashed black box indicates a kagome bilayer. (b) The kagome bilayer displayed in the $ab$ plane.\label{fig:struct_bilayer}}
\end{figure}

To summarize we have solved the crystal structure of Ca$_{10}$Cr$_7$O$_{28}$ and shown that it can be well described by the conventional unit cell where the Cr3 and O3 sites are both disordered between two positions (Cr3A/Cr3B and O3A/O3B) whose occupancies have the ratio 3:1 and sum to 100\%. A more realistic model is for the Cr3A/Cr3B and O3A/O3B sites to order giving rise to a doubling of the $a$ and $b$ lattice parameters while preserving all other aspects of the crystal structure. Evidence for this supercell comes from the observation of superlattice peaks at half-order positions as we have found in the X-ray single crystal data. Ca$_{10}$Cr$_7$O$_{28}$ is a mixed valence compound with a ratio of Cr$^{6+}$ to Cr$^{5+}$ of 6:1. By performing a bond valence sum we showed that the Cr3A and Cr3B ions have valence +6 while the Cr1 and Cr2 ions have valence +5. This charge order has important consequences for the magnetism because while Cr$^{6+}$ ions are non-magnetic, Cr$^{5+}$ ions have a spin value of S=1/2. Magnetization and susceptibility measurements confirm that 6 out of 7 Cr ions have spin-1/2 in accordance with the number ratio of Cr1 and Cr2 ions to Cr3A and Cr3B ions.

From the point of view of the magnetic behavior, the Cr3A and Cr3B are unimportant and can be neglected. It should be further mentioned that there is no difference concerning the magnetic Cr1 and Cr2 ions between the convention unit cell and the supercell since the superstructure results from the site ordering of the non-magnetic Cr3A/Cr3B ions and the O3A/O3B ions while the other ions are unaffected. The positions of the magnetic Cr1 and Cr2 ions within the convention unit cell along with the seven nearest neighbor exchange interactions that couple them are shown in Fig.~\ref{fig:struct_bilayer}. They form a kagome bilayer structure within the $ab$-plane where the bilayers are stacked in an ABC type arrangement along the $c$ axis. Both kagome layers that form the bilayers consist of two equilateral triangles of different sizes (and therefore inequivalent interactions) that share corners and alternate throughout the plane. This particular arrangement of corner-sharing spin$-\frac{1}{2}$ triangles coupled into bilayers has not been reported in any other compound. It provides the source of highly frustrated magnetism that leads to the quantum spin liquid behavior observed in Ca$_{10}$Cr$_7$O$_{28}$ as described by C. Balz \emph{et al.} in Ref.~\cite{Bal16}.

\begin{acknowledgments}
\emph{Acknowledgments} - The authors acknowledge useful discussions with Thomas C. Hansen and thank Bachir Ouladdiaf for access to the OrientExpress instrument.
\end{acknowledgments}

\appendix*
\section{Appendix: Refinement}

\begin{turnpage}
\begin{table}
\caption{Result of the crystal structure refinement of Ca$_{10}$Cr$_7$O$_{28}$ at 300\,K using the best model (\emph{model 3}) described in the text. The refinement was carried out in the tetragonal space group $R3c$. The lattice parameters were obtained from the X-ray synchrotron powder refinement and used in the neutron single crystal refinement. The fractional $x$-, $y$- and $z$-parameters were taken from the single crystal refinement and used in the X-ray powder refinement. The anisotropic thermal parameters $B_{ij}$[\AA$^2$] were obtained from the single crystal data. They are multiplied by $\times10^{4}$ to improve legibility. For symmetry reasons $B11$ and $B22$ of the atoms Ca4, Cr3, and O3 are identical. Furthermore, for these atoms $B13$ and $B23$ are equal to zero. The isotropic thermal parameters are from the X-ray synchrotron data. The occupancies of Cr3A, Cr3B, O3A and O3B were refined from the single crystal data and found to be 0.726(12) for Cr3A and 0.276(12) for Cr3B. The same ratio applies to O3A and O3B.\label{tab:refinement}}
\begin{ruledtabular}
\begin{tabular}{c c c c c c c c c c c}
\multicolumn{2}{l}{Lattice Parameters} \\
\hline
$a$ & 10.76892(3)\AA\\
$c$ & 38.09646(12)\AA\\
\hline
Atom & $x$ & $y$ & $z$ & $B_{11}(\times10^{4})$ & $B_{22}(\times10^{4})$ & $B_{33}(\times10^{4})$ & $B_{12}(\times10^{4})$ & $B_{13}(\times10^{4})$ & $B_{23}(\times10^{4})$ & $B_{iso}$[\AA$^2$]\\
\hline
Ca1 & 0.28548(17) & 0.15745(16) & -0.05744(3) &71.4(1.6) &   49.3(1.6) &    2.3(0.1) &   41.2(1.4) &   -5.5(0.3) &   -5.3(0.3)& 1.661(38)\\
Ca2 & 0.19737(17) & -0.19624(18) & 0.00371(0) &23.4(1.2)    &29.5(1.4)     &1.4(0)&    11.0(1.0)&    -0.4(0.2)&    -1.4(0.2)& 0.754(18)\\
Ca3 & 0.38806(16) & 0.18233(17) & 0.03747(2) &27.7(1.3)&    26.8(1.3)&     1.4(0)&    11.7(1.1)&    -1.7(0.2)&    -0.8(0.2)& 0.754(18)\\
Ca4 & 2/3 & 1/3 &  0.10343(7) &41.1(1.5)&    41.1(1.5)&     1.0(0.1)&    20.6(0.7)&     0&     0& 0.794(42)\\
\hline
Cr1&0.31133(19)&0.14053(21) & 0.13588(4)& 22.0(1.5)&    30.8(1.8)&     1.1(0.1)&    15.8(1.3)&     0.7(0.2)&    -0.1(0.3)& 0.564(12)\\
Cr2&0.17771(18) & -0.13669(19) & -0.09600(4)& 24.8(1.5)&    22.9(1.6)&     1.0(0.1)&     15.6(1.3)&     -0.1(0.2)&     0.4(0.2)& 0.564(12)\\
Cr3A & 0 & 0 &  0.00446(10)& 23.0(3.7)&    23.0(3.7)&     2.0(0.2)&    11.5(1.8)&     0&     0& 0.564(12)\\
Cr3B & 0 & 0 & 0.02961(23)& 25.9(9.9)&    25.9(9.9)&     1.7(0.5)&    12.9(4.9)&     0&     0& 0.564(12)\\
\hline
O3A & 0 & 0 &  -0.03863(9)& 79.4(4.7)&    79.4(4.7)&     2.8(0.2)&    39.7(2.4)&     0&     0&  2.254(55)\\
O3B & 0 & 0 &  0.07192(26)& 105.0(15.2)&    105.0(15.2)&     2.7(0.5)&    52.5(7.6)&     0&     0&  2.254(55)\\
O1 & 0.28286(18) & 0.08966(16) & 0.09384(3)& 88.8(1.8)&    73.3(1.7)&     1.1(0)&    47.7(1.5)&    -0.6(0.2)&     -0.1(0.2)& 2.254(55)\\
O2 & 0.23570(17) & 0.23520(18) & 0.14947(3)& 57.1(1.4)&    65.5(1.5)&     2.2(0)&    50.0(1.3)&     3.2(0.2)&     1.7(0.2)& 2.254(55)\\
O3 & 0.27626(12) & -0.01602(12) & 0.15721(3)& 26.2(1.0)&    23.6(0.9)&     1.5(0.0)&     6.2(0.8)&     0.8(0.2)&     0.6(0.2)& 0.872(49)\\
O4 & 0.49099(14) & 0.24584(15) & 0.14533(3)& 19.9(1.0)&    21.7(1.1)&     3.1(0.1)&     6.9(0.9)&     1.1(0.2)&     0.3(0.2)& 0.872(49)\\
O5 & 0.14995(20) & -0.12682(19) & -0.05274(3)& 84.6(1.9)&    77.2(1.8)&     0.9(0)&    12.9(1.5)&    0.5(0.2)&    -1.0(0.2)& 2.254(55)\\
O6 & 0.25594(15) & -0.23730(15) & -0.10634(4)& 29.8(1.1)&    27.7(1.1)&     2.9(0.1)&    19.6(1.0)&     0.6(0.2)&    0.9(0.2)&  0.872(49)\\
O7 & 0.29992(15) & 0.03281(12) & -0.10850(4)& 52.9(1.2)&    25.5(1.0)&     1.7(0)&    16.8(0.9)&     0.7(0.2)&     1.1(0.2)& 0.872(49)\\
O8 & 0.01423(15) & -0.22258(1) & -0.11430(4)& 26.9(1.2)&    93.2(2.0)&     2.2(0.1)&     20.6(1.3)&     -2.8(0.2)&    1.1(0.3)& 2.254(55)\\
O9 & -0.00401(22) & 0.14481(14) & 0.01693(5)& 49.8(1.1)&    32.8(1.3)&     4.6(0.1)&    26.2(1.2)&     6.1(0.3)&     2.5(0.3)& 2.254(55)\\
\end{tabular}
\end{ruledtabular}
\end{table}
\end{turnpage}

\end{document}